\documentclass[aps,prb,reprint,superscriptaddress]{revtex4-1}

\usepackage{graphicx}
\usepackage{amsmath}
\usepackage{textcomp} 
\usepackage{xcolor}
\usepackage{ulem}
\usepackage{amssymb}

\begin{document}


\title{Control of spin dynamics in artificial honeycomb spin-ice-based nanodisks}%
\author{Mojtaba Taghipour Kaffash}
\affiliation{Department of Physics and Astronomy, University of Delaware, Newark, DE 19716, USA}
\author{Wonbae Bang}
\affiliation{Institute of Advanced Materials, LG Chem, Daejeon 34122, Korea}
\affiliation{Department of Physics and Astronomy, Northwestern University, Evanston, IL 60208, USA}
\affiliation{Materials Science Division, Argonne National Laboratory, Argonne, IL 60439, USA} 
\author{Sergi Lendinez}
\affiliation{Department of Physics and Astronomy, University of Delaware, Newark, DE 19716, USA}
\author{Axel Hoffmann}
\affiliation{Materials Science Division, Argonne National Laboratory, Argonne, IL 60439, USA} 
\affiliation{Department of Materials Science and Engineering, University of Illinois at Urbana-Champaign, Urbana, IL 61801, USA.}

\author{John B. Ketterson}
\affiliation{Department of Physics and Astronomy, Northwestern University, Evanston, IL 60208, USA}
\author{M. Benjamin Jungfleisch}
\email{mbj@udel.edu}
\affiliation{Department of Physics and Astronomy, University of Delaware, Newark, DE 19716, USA}



\date{\today}

\begin{abstract}
We report the experimental and theoretical characterization of the angular-dependent spin dynamics in arrays of ferromagnetic nanodisks arranged on a honeycomb lattice.
The magnetic field and microwave frequency dependence, measured by broadband ferromagnetic resonance, reveal a rich spectrum of modes that is strongly affected by the microstate of the network.
Based on symmetry arguments with respect to the external field, we show that certain parts of the ferromagnetic network contribute to the detected signal. A comparison of the experimental data with micromagnetic simulations reveals that different subsections of the lattice predominantly contribute to the high-frequency response of the array. This is confirmed by optical characterizations using microfocused Brillouin light scattering. Furthermore, we find indications that nucleation and annihilation of vortex-like magnetization configurations in the low-field range affect the dynamics, which is different from clusters of ferromagnetic nanoellipses. Our work opens up new perspectives for designing magnonic devices that combine geometric frustration in gyrotropic vortex crystals at low frequencies with magnonic crystals at high frequencies.
\end{abstract}

\maketitle


\section{\label{sec:level1}Introduction}
The investigation of artificial spin ice (ASI) 
structures has received increased attention in the magnetism community over the past decade \cite{Nisoli_RMP2013,Gilbert_Physics_Today,Heyderman_JPCM2013,Lendinez:2019hk,iacocca2019,Skjaervo_2019,ortizambriz_2019}. Beyond the traditional studies on slow dynamics and thermalization effects in ASI, there is an effort to understand the spin dynamics in the microwave frequency regime in those networks, e.g., Refs.~[\onlinecite{Zhou_Adv2016,Bang_2019,Bhat_PRB_2018,Bhat_PRB2016,Iacocca_PRB2016,Jungfleisch_PRB_Ice_2016,Gliga_PRL2013,Branford_PRB_2019,Branford_PRB_2019_2}] as well as their magnetotransport properties, e.g., Refs.~[\onlinecite{Tanaka_PRB_2006,Branford_Science_2012,Jungfleisch:2017fb,Le_PRB2017,Jungsik_PRB}]. Aided by advances in modern nanofabrication technologies it is possible to create two-dimensional arrays of ferromagnetic nanoscaled elements with a wealth of possible orientations and alignments, and to place these elements on a lattice. The original intention of creating ASI was to design geometrically frustrated networks that closely mimic the frustration in crystalline spin ices, e.g., Refs.~[\onlinecite{Wang_Nature_2006,Bramwell_Science_2001,Morris_2009,Fennell_2009}]. Many different lattice structures such as honeycomb, Shakti, Tetris, brickwork, etc. have been investigated since then \cite{park2017magnetic,stopfel2018magnetic,zhang2013crystallites,gilbert2016emergent}. What these lattices have in common is that the ferromagnetic elements consist of elongated singe-domains, also referred to as `\textit{islands}', in which shape anisotropy facilitates an alignment of the magnetic moments to point parallel or antiparallel to the easy axis of the island. 
As such, these elements essentially have only binary degrees of freedom. Recently, new types of non-Ising like lattices such as XY spin systems \cite{Streubel_2018} and Potts ASI \cite{Louis_2018,Sklenar_2019} have been explored and more exotic phase transitions were found. 

\begin{figure}[b]
\centering
\includegraphics[width=.5\textwidth]{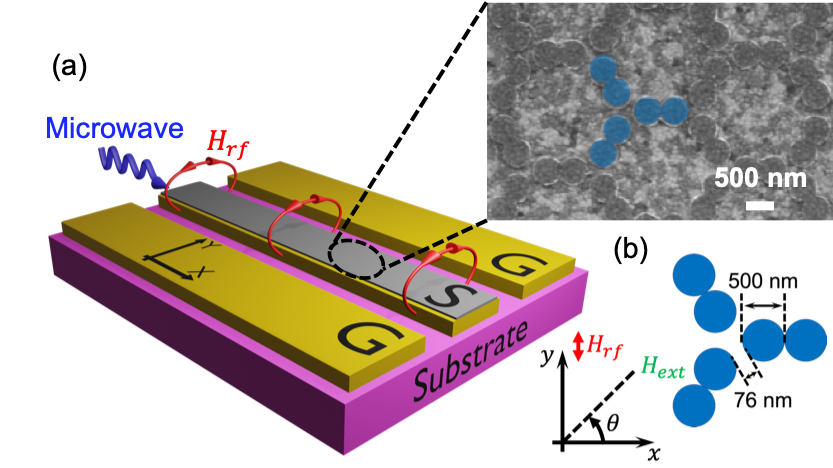}
\caption{(a) Schematic view of the experimental FMR setup including SiO$_2$ substrate (purple), coplanar waveguide (CPW) (yellow) with two ground lines (G) and one signal line (S), and artificial spin ice sample (gray) patterned on top of the CPW. The inset shows a scanning electron micrograph of the fabricated honeycomb lattice. (b) Schematic view and dimensions of a vertex of the honeycomb lattice with the direction of the microwave magnetic field $H_\mathrm{rf}$ and the direction of the in-plane angle $\theta$ of the external magnetic field $H_\mathrm{ext}$.} 
\label{fig 1}
\end{figure}

From a dynamic perspective, recent work by Behncke et al. showed that arrays of ferromagnetic disks are particularly interesting as their polarization state can be tuned so that geometrical frustration arises through the presence and motion of magnetic vortices in the disks \cite{behncke2018tunable}. They showed that frustrated and non-frustrated states can be achieved by changing the frequency of the state formation process. However, the individual disks had micrometer diameters -- too big to be treated as a XY spin system -- and only the low-frequency dynamics, corresponding to the gyrotropic frequencies, of the vortices were studied.

Here, we describe detailed experimental and theoretical characterizations of  angular-dependent spin dynamics observed in arrays of ferromagnetic nanodisks. Our nanodisks have a diameter well below one micrometer and are closely packed on a honeycomb lattice. The broadband ferromagnetic resonance (1 -- 10 GHz) measurements are complemented by micromagnetic simulations using mumax$^3$ [\onlinecite{mumax3}] and micro-focused Brillouin light scattering ($\mu$-BLS) spectroscopy. Our results show that vortex-like magnetization configurations in the disks affect the high-frequency dynamics in the honeycomb-disk ASI networks, which differs from clusters of ferromagnetic nanoellipses \cite{Bang_JAP2019}. Furthermore, we find evidence of localized magnetization dynamics in our arrays.
This is the first step in combining geometric frustration in a gyrotropic vortex crystal at low frequencies with a magnonic crystal at high frequencies.






\section{Experimental details and micromagnetic approach}
In the following, we present details on the sample fabrication, ferromagnetic resonance measurements, micromagnetic simulations, as well as micro-focused Brillouin light scattering spectroscopy.

\subsection{Sample fabrication}
A multi-step lithography process was used for sample fabrication. In the first step, a coplanar waveguide (CPW) was defined on a thermally oxidized Si substrate using optical lithography on to which 5-nm Ti and 120-nm of Au was deposited by electron-beam evaporation followed by lift-off. 
The width of the signal line is 20 $\mu$m separated by an 8 $\mu$m gap between the signal and the 40 $\mu$m wide ground lines. We have previously shown that the best coupling between the microwave magnetic field created by the CPW and arrays of nanomagnets is obtained by directly patterning the arrays on top of the signal line \cite{Jungfleisch_PRB_Ice_2016}. Hence, the ASI structure was written on the signal line by electron-beam lithography. For this purpose, a double layer positive resist of MMA (methyl methacrylate)/PMMA (polymethyl methacrylate) was used. After exposure and development of the resist stack, we used electron-beam evaporation to deposit 
15-nm of permalloy (Py, Ni$_{80}$Fe$_{20}$), followed by a lift-off step. A sketch of the fabricated CPW with the magnetic material is shown in Fig.~\ref{fig 1}(a). A scanning electron microscopy image of the honeycomb lattice made of ferromagnetic nanodisks on top of the signal line is shown as an inset. Figure~\ref{fig 1}(b) shows the dimensions of the disks as well as the separation of the disks to neighboring disks in the honeycomb lattice.  The disks have a diameter of 500 nm and are constructed as a pair, where the gap between neighboring pairs is 76 nm. 


\subsection{Broad-band ferromagnetic resonance (FMR) measurements}
The patterned devices were characterized by broadband ferromagnetic resonance using a vector network analyzer (VNA) \cite{Kalarickal_2006}. An external magnetic field $H_\mathrm{ext}$ is applied in the sample plane, and its angle $\theta$ with respect to the CPW can be controlled by an automated rotating motor, see Fig.~\ref{fig 1}(b). As shown in Fig.~\ref{fig 1}(a), a microwave current is passed through the CPW, generating an alternating magnetic field H$_\mathrm{rf}$ in the $y$-axis direction (red arrows). H$_\mathrm{rf}$ excites a collective precessional motion of the magnetic moments in the disks, which can be detected using the VNA. 

The FMR spectra were recorded as follows: the sample was first saturated by an external in-plane magnetic field at $-2100$ Oe, followed by recording a reference spectrum at $-1500$~Oe. Thereafter, the field was gradually increased from $-1000$~Oe to 1000~Oe in steps of 10~Oe while sweeping from 1~GHz to 10~GHz and recording the transmission S21 parameter using the VNA at each field step. 
The reference spectrum is then subtracted from the spectra at each field step to account for changes in the transmission characteristics not associated with the magnetic field. 

\subsection{Micromagnetic simulations}
\label{simulations}
All simulations were conducted using the mumax$^3$ micromagnetic simulator \cite{mumax3}. For this purpose, each of the 15~nm thick 500~nm diameter nanodisks is partitioned into $5\times5\times15$ nm$^3$ 
so that the lateral dimensions are kept less than the exchange length ($L_\mathrm{ex}^\mathrm{Py}=5.3$ nm). The grid is divided in $512\times512\times1$ cells. Standard magnetic parameters for Py are used: saturation magnetization $M_\mathrm{sat}=800\times 10^3$ A/m, exchange stiffness $A_\mathrm{ex}=13\times 10^{-12}$ J/m and Gilbert damping parameter $\alpha=0.01$. A honeycomb vertex is formed of 6 disks as shown in Fig.~\ref{fig 1}. Two disks build a pair that is arranged on a honeycomb lattice, where the connecting axis between  two disks is aligned along the signal line of the CPW, while the axes of the two neighboring coupled disks are directed at an angle $\theta=\pm 120^{\circ}$, respectively.


\begin{figure*}[t]
\centering
\includegraphics[width=1\textwidth]{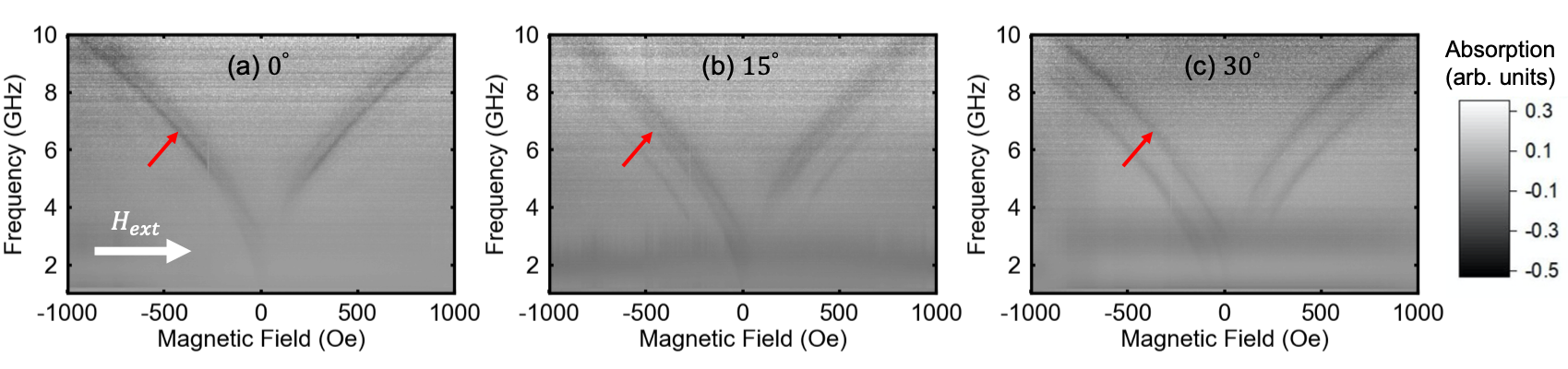}
\caption{Experimental absorption spectra of our sample, obtained by angular-dependent broadband ferromagnetic resonance. The external magnetic field $H_\mathrm{ext}$ is applied at different in-plane angles (a) $\theta=0^{\circ}$, (b) $\theta=15^{\circ}$, (c) $\theta=30^{\circ}$ with respect to the signal line of the CPW ($x$-axis). As the magnetic field is swept from negative to positive values, the spectra are recorded by a vector network analyzer at each field step. The white arrow shows the magnetic field sweep direction. The red arrow indicates the main mode as discussed in the text.}
\label{fig 2}
\end{figure*}

\subsection{Microfocused Brillouin light scattering ($\mu$-BLS)}
The spatially resolved dynamic response of the resonant modes in the lattice was measured using $\mu$-BLS \cite{Sebastian_2015}. For this purpose, we use a 532~nm single mode laser with a continuous power of less than 2~mW at the sample position. The external field is applied parallel to the direction of the signal line of the CPW ($\theta = 0^{\circ}$). The same sample was used for both the $\mu$-BLS and the FMR measurements. Furthermore, we use the same field routine as in the FMR measurements and micromagentic simulations: First, the sample is saturated and then the field is swept, while changing the excitation frequencies and probing the excited dynamics. A nominal microwave power of +22~dBm is used, which is low enough to avoid any non-linearities. Informed by the acquired field/frequency response, we choose a particular field/frequency combination to measure the spatial extent of the spin dynamics.


\section{Results and discussion}

Figures~\ref{fig 2}(a-c) show the experimental FMR spectra as false color-coded images for the nanodisks arranged on a honeycomb lattice for three different in-plane angles $\theta$ of $0^\circ$, $15^\circ$ and $30^\circ$ with respect to the signal line of the CPW ($x$-axis). Note that the magnetic field is swept from negative to positive fields. A dark contrast shows a strong microwave absorption indicative of an efficient excitation of spin dynamics in the array, while a brighter color means a negligible microwave absorption (i.e., an absence of coherent spin dynamics).
The choice of these three external field angles is based on the symmetry of the lattice and the fact that the angles $45^{\circ}$, $60^{\circ}$, $75^{\circ}$ and $90^{\circ}$ show similar behavior as the former three. In other words, by shape symmetry, one can infer that there is a similarity between the following groups of external applied field angles: i) $0^{\circ}$ and $60^{\circ}$; ii) $15^{\circ}$, $45^{\circ}$ and $75^{\circ}$; and iii) $30^{\circ}$ and $90^{\circ}$. 



As is shown in Fig.~\ref{fig 2}, we detect a number of different modes in the FMR spectra 
with varying position and intensity depending on the in-plane magnetic field angle $\theta$. The spectra can be divided into three different regimes: One at high fields, one at low fields, and one at intermediate fields. 

The first regime ranges from $-1000$~Oe to 20~Oe when sweeping the magnetic field up, i.e., it starts with a configuration in which all magnetic moments in the disks are saturated along $-H_\mathrm{ext}$ and ends just short of an instability. 

In this region, the mode indicated by the red arrow in Fig.~\ref{fig 2} corresponds to the bulk-like fundamental mode, and is produced by the spins that are mostly aligned along the external magnetic field. It shows the strongest absorption for the applied field at $\theta=0^{\circ}$. As the angle $\theta$ is increased, the intensity of this mode decreases, while other lower-lying modes appear in the spectra, see Figs.~\ref{fig 2}(b) and (c). Furthermore, the fundamental mode is shifted slightly to higher frequencies as the angle $\theta$ is increased. 

\begin{figure*}[t]
\centering
\includegraphics[width=1\textwidth]{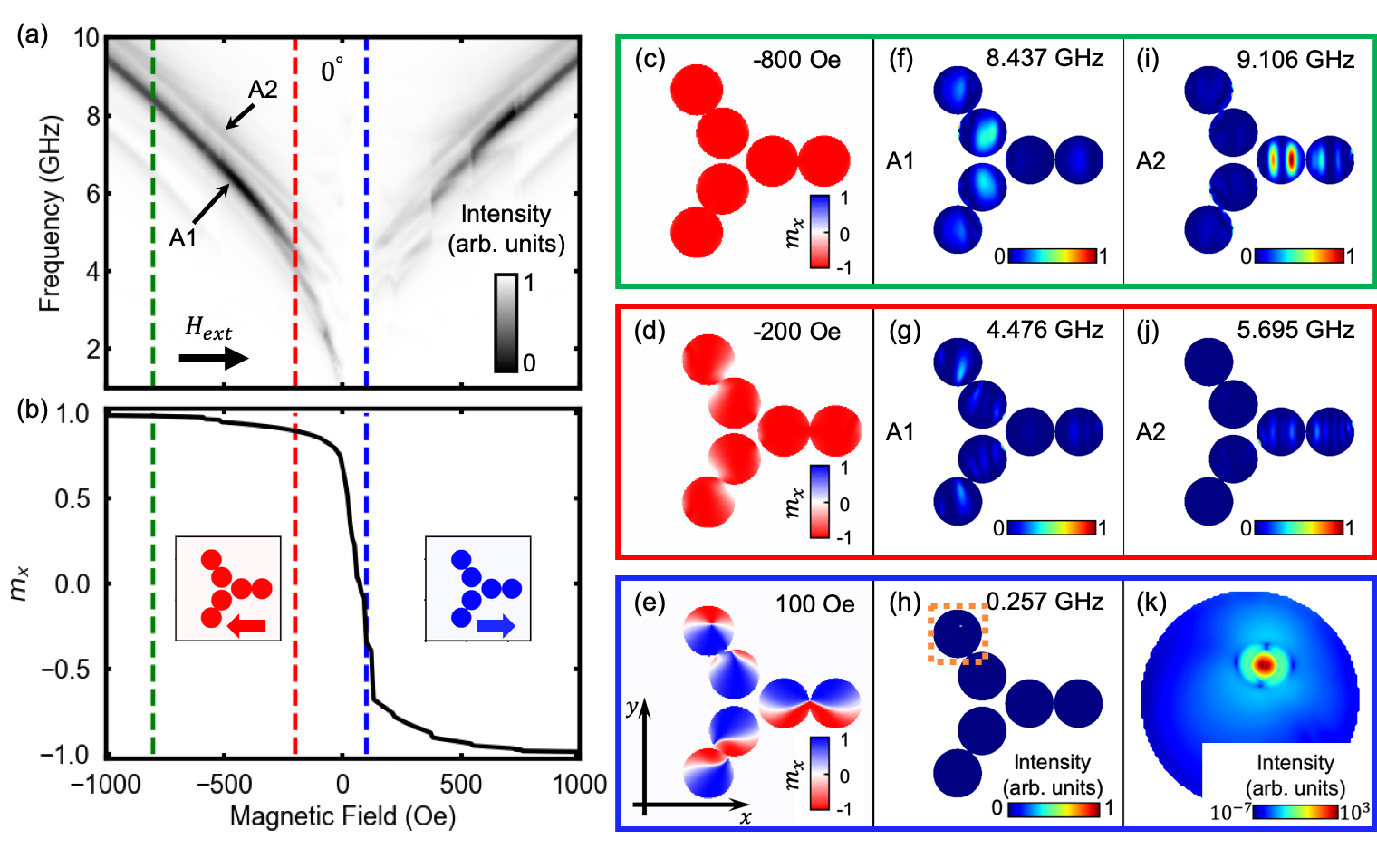}
\caption{Results of micromagnetic simulations. (a) Simulated absorption spectra for an applied in-plane filed at $\theta=0^{\circ}$. The black arrow indicates the direction in which the field is swept. (b) Corresponding magnetization behavior of the normalized $x$-component of the magnetization ($m_x$). The insets shows the saturated configuration of $m_x$ in one vertex for $-1000$~Oe (red) and 1000~Oe (blue). (c)-(e) Normalized $m_x$-component value at fields $-800$~Oe (green), $-200$~Oe (red), and 100~Oe (blue), respectively. These field are labeled with dashed lines in (a) and (b). (f)-(k) Corresponding 2D FFT intensity profiles for frequencies with strong absorption in both experimental and simulation results. (k) Magnified vortex dynamics for the $120^\circ$ outer disk in (h) indicated be dashed square. Note that only (k) is plotted in logarithmic scale; all other figures are in linear scale. 
} 
\label{fig 3}
\end{figure*}

In the second regime, ranging from  $20$ to $120$~Oe, we observe a distinct \textit{gap} in the mode spectrum independent of the in-plane field angle. Typically, one can observe a decrease of the resonant frequency in this field range, followed by the nucleation of magnetic vortices in the disks accompanied by the onset of a gyrotropic motion of the vortices \cite{behncke2018tunable}. The gyrotropic motion typically occurs at frequencies well below 1 GHz (depending on the dimensions of the disks and specific material parameters). While we do observe the gyrotropic motion in micromagnetic simulations (as it will be discussed bellow), we did not detect any signal below 1 GHz in our experiments. There are two reasons for this, both of which arise from the smaller thickness and lateral dimensions of our sample compared to the honeycomb disks studied in Ref.~[\onlinecite{behncke2018tunable}].  First, since vortex dynamics generally produce a weak signal, less thickness can make the signal detection more challenging. Second, as explained in detail below, the simulated magnetization configuration [Fig. \ref{fig 3} (e)] suggests that only in one nanodisk (out of six) at each honeycomb vertex is a vortex formed. Even more importantly, the magnetic moments in the disks curl, and thus the net magnetization in each disk is almost zero. As a result, the overall FMR response is minimum as is evident from the gap in the Fig.~\ref{fig 2}. It is interesting to note that this is independent of the in-plane field angle, which is different from clusters of ferromagnetic nanoellipses \cite{Bang_JAP2019}.

The third regime ranges from $120$ to $300$~Oe. In this field range not all moments have aligned with the external field direction. This leads to some minor differences in the resonant dynamics compared with the symmetric negative fields. These differences are absent in the high-field regime.

The last part of the spectra ($300$ to $1000$~Oe) shows the same behavior as the negative field regime at high magnetic field values. Here, all magnetic domains are completely aligned with $H_\mathrm{ext}$ and the resonant spectra are similar to the negative field range.

\begin{figure*}[t]
\centering
\includegraphics[width=1.0\textwidth]{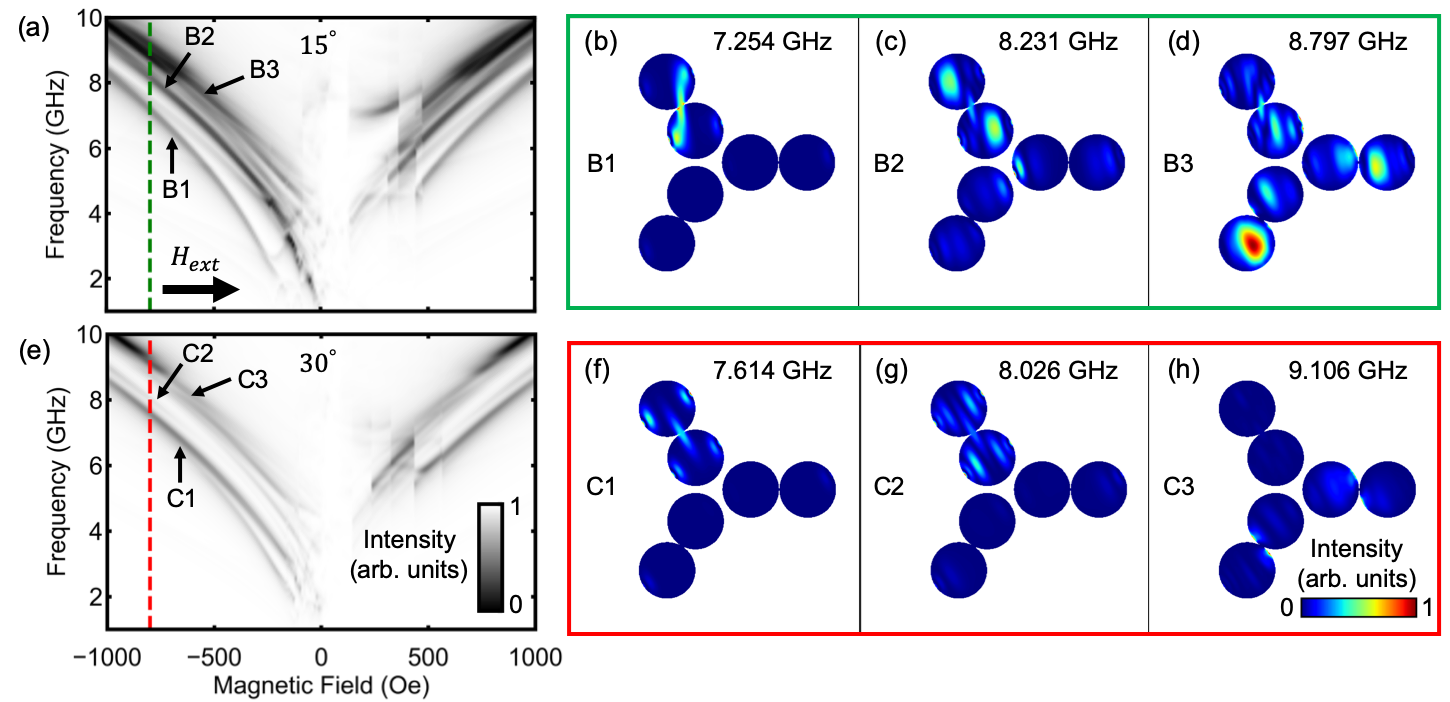}
\caption{Results of micromagentic simulations. (a) Simulated absorption spectra for an applied in-plane field angle of $\theta=15^{\circ}$ in a color-coded image, and (b)-(c) the spatial profiles of the simulated spin dynamics at $-800$~Oe [green dashed line in (a)] and frequencies of (b) 7.254~GHz, (c) 8.231~GHz, and (d) 8.797 GHz. (e) Simulated absorption spectra for an applied in-plane field angle of $\theta=30^{\circ}$, and (f)-(h) the spatial profiles of the spin dynamics at $-800$~Oe [red dashed line in (e)] and frequencies of (f) 7.614~GHz, (g) 8.026~GHz, and (h) 9.106 GHz.}
\label{fig 5}
\end{figure*}



The experimental FMR spectra are the result of the collective dynamics of the lattice. However, in order to get more insight of the two-dimensional profiles of the magnetization configuration and the spin dynamics we perform micromagnetic simulations.
For this purpose, a procedure replicating the experimental routine was implemented in mumax$^3$: First, the magnetization of the lattice is saturated by initializing all the spins in $-H_\mathrm{ext}$ direction. 
 Then, the field is swept from $-1000$~Oe to 1000~Oe [indicated by a black arrow in Fig.~\ref{fig 3}(a)], with 10~Oe increment steps. The equilibrium magnetization state at each field step is obtained by minimizing the total energy and then letting the magnetization evolve for 5~ns. This is done to avoid possible saddle points in the energy landscape. In order to obtain the frequency response, a perturbative sinc-shaped field pulse with an amplitude of 1~Oe and cut-off frequency of 50~GHz is applied in the $y$-axis. The time evolution of the magnetization is recorded every 1~ps for a total time of 20~ns. A fast-Fourier transform (FFT) is performed on the $z$-component of the time-dependent magnetization to obtain the characteristic precession frequencies of the system at each field step. False-color coded plots of the FFT intensity as a function of the external magnetic field and frequency can then be obtained, as shown in Fig.~\ref{fig 3}(a).


By comparing the experimental FMR data [Figs.~\ref{fig 2}(a-c)] with their corresponding simulation results [Fig. \ref{fig 3}(a)] we find a very good agreement for the different field regimes discussed above. As we will see, that is also true for the other in-plane field angles studied here [Figs. \ref{fig 5}(a) and (e)]. Before we turn to a discussion of the spin dynamics, we focus on the static magnetization configuration. Figure~\ref{fig 3}(b) shows the simulated field scan of the normalized $x$-component of the static magnetization ($m_x$). The vertical dashed lines indicate the same field values as in Fig.~\ref{fig 3}(a). They denote field values of $-800$~Oe (green), $-200$~Oe (red), and +100~Oe (blue). The two insets in Fig.~\ref{fig 3}(b) depict the color code for the normalized magnetization $m_x$, displaying the disks of one honeycomb vertex in the saturated state, i.e., at $-1000$~Oe (red) and at $+1000$~Oe (blue). The magnetization configuration of one honeycomb vertex is illustrated in Figs.~\ref{fig 3}(c)-(e) at the intermediate fields as indicated by the dashed lines in (a) and (b) [$-800$~Oe (green), $-200$~Oe (red), and +100~Oe (blue)].

As is obvious from the simulated magnetization vs. external field ($m_\mathrm{x}$ vs. $H_\mathrm{ext}$) trace shown in Fig.~\ref{fig 3}(b), the magnetization configuration points mostly along the field direction in the high-field region. As the field is approaching zero, the magnetic moments in the nanodisks start to flip as can been seen from the steps in the $m_\mathrm{x}$ vs. $H_\mathrm{ext}$ plot [Fig.~\ref{fig 3}(b)], and from the spatially-resolved simulation results: Fig.~\ref{fig 3}(c) at $-800$~Oe, Fig.~\ref{fig 3}(d) at $-200$~Oe, and Fig.~\ref{fig 3}(e) at 100~Oe. This field range, in which the magnetization configuration changes corresponds to the \textit{gap} region observed in the dynamics measurements [Fig.~\ref{fig 2}(a)] and simulations [Fig.~\ref{fig 3}(a)], where no resonant dynamics are observed. Increasing the magnetic field even further results in a progressive re-alignment of the moments to the positive field direction until the overall magnetization points in the positive direction.

In addition to the FFT intensity spectra, two-dimensional profiles of the FFT intensity were computed at selected magnetic fields/frequencies. First, we discuss the two-dimensional profiles for an in-plane magnetic field angle of 0$^\circ$. The corresponding 2D maps are shown in Figs.~\ref{fig 3}(f)--(k). As mentioned above Figs.~\ref{fig 3}(c)--(e) show the magnetization configuration, while (f)--(k) depict the spatially-resolved dynamics in the elements. For this purpose, the average magnetization as a function of time as well as the magnetic configuration at each time step were computed. The dynamic profiles were obtained by calculating the FFT of each cell. The following discussion focuses on three particular magnetic field values, $-800$~Oe, $-200$~Oe, and 100 Oe, as indicated by the dashed lines (green, red, and blue, respectively) in Figs.~\ref{fig 3}(a) and (b).

At $-800$ Oe, the first fundamental mode [mode A1 in Fig. \ref{fig 3}(a)] with one antinode in an individual disk (frequency 8.437~GHz) is localized in the $\pm 120^{\circ}$ elements [Fig. \ref{fig 3}(f)]. This mode has the strongest intensity in the spectra. On the other hand, the second prominent mode indicated in Fig.~\ref{fig 3}(a) as mode A2 is a higher-order mode localized in the horizontal nanodisks, see Fig.~\ref{fig 3}(i). This mode is less intense and lies at a higher frequency than mode A1. It is characterized by two antinodes.

As the field is decreased, the resonant frequency gradually decreases. The spatially resolved dynamic profiles at $-200$~Oe are shown in Figs.~\ref{fig 3}(g) and (j). As is visible in the 2D maps, higher-order modes with three anti\-nodes contribute to the signal at 4.476~GHz [Fig. \ref{fig 3}(g)] and five antinodes to the signal at 5.695~GHz [Fig. \ref{fig 3}(j)] in the lower frequency regime. Note that the difference in intensity in each pair of disks is due to the fact that the vertex was simulated without taking into account the neighboring disks to reduce the computation time. We tested simulations with neighboring disks, i.e. a larger array (not shown here), to confirm that the magnetostatic interactions from the other disks cause the intensity to be equally distributed between the pairs.

In a low field of 100~Oe 
the magnetization configuration further creates flux closure domains to minimize the magnetostatic energy, see Fig.~\ref{fig 3}(e). This lowers the internal effective field and, hence, the resonance frequency drops to MHz frequencies. Moreover, the only disk that experiences the vortex state is the outer $120^{\circ}$ element. As seen from the 2D profile of the FFT intensity, Fig.~\ref{fig 3}(h), we observe the vortex gyromotion in the outer disk at $120^\circ$ at 257 MHz; a magnified 2D plot of the vortex dynamics is shown in Fig.~\ref{fig 3}(k) in logarithmic scale.
Although the geometry of the structure is symmetric with respect to the $x$-axis, the magnetic configuration is not, as can be seen from Fig.~\ref{fig 3}(e). This is because in our simulations the external field $H_\mathrm{ext}$ is applied at $1^{\circ}$ with respect to the $x$-axis to break the symmetry and, thus, to avoid possible saddle points in the simulations. 
Depending on the initial conditions (e.g. field sweep direction, the choice of a small deviation angle for the magnetic field, etc.), the vortex state can be created in different nanodisks. However, in our simulations it was not possible to create the vortex state in all nanodisks of the lattice at the same time. This could also be the reason for us not being able to experimentally detect the corresponding mode.

The high-field regime at positive fields (i.e. $+150 - +1000$~Oe) resembles the behavior at negative fields and, thus, we omit a detailed discussion here.

To demonstrate the ability to turn on the precession in different parts of the network, we compare the experimental FMR results at different in-plane angles with micromagnetic simulations. 
Figure~\ref{fig 5}(a) presents the simulated frequency versus magnetic field spectra for $\theta = 15^\circ$, while Fig.~\ref{fig 5}(b) shows the corresponding simulations for $\theta = 30^\circ$. We first discuss the $15^\circ$ data. In particular, we focus on the modes labeled as B1, B2 and B3 in Fig.~\ref{fig 5}(a). Their corresponding spatial profiles at $- 800$~Oe [indicated by a vertical green dashed line in (a)] are shown in Figs.~\ref{fig 5}(b)-(d).

Due to the applied field direction, the axis of symmetry for the modes is at $120^{\circ}$ (compared with the $0^{\circ}$ direction of $H_\mathrm{ext}$ when it lies along the $0^{\circ}$ axis). Mode B1, which has the lowest frequency, is localized in the $120^{\circ}$ disks. As is obvious from Fig.~\ref{fig 5}(b), mode B1 arises from a coupling of the resonances in the pair of disks aligned at $120^{\circ}$. Mode B2 lies between B1 and B3 in Fig.~\ref{fig 5}(a). In this intermediate frequency range, the resonant response is distributed in all nanodisks, while the strongest precession is found in the $120^{\circ}$ disks [Fig.~\ref{fig 5}(c)]. The strongest mode at $-800$~Oe is observed in the simulations for mode B3, which is in fact a band of modes. The resonance in the spectrum [Fig.~\ref{fig 5}(a)] mainly stems from the $0^{\circ}$ and $-120^{\circ}$ disks as indicated in Fig.~\ref{fig 5}(d).

Figures~\ref{fig 5}(e)-(h) show the simulation results when the field $H_\mathrm{ext}$ is applied at $30^{\circ}$ with respect to the $x$-axis. At a large field [red dashed line for $-800$~Oe in Fig. \ref{fig 5}(e)], the simulated spectrum shows two intense modes (C1 and C3) that are well separated by a weaker mode (labeled as C2). The corresponding spatial profiles are shown in Figs.~\ref{fig 5}(f)-(h). Modes C1 and C2 are localized in the $120^{\circ}$ disks. Again, the modes are coupled along the axes of symmetry of the corresponding pairs of disks [Figs.~\ref{fig 5}(f) and (g)], whereas mode C3 is localized in the $0^{\circ}$ and $-120^{\circ}$ elements. This result demonstrates the possibility of controlling the dynamics in different parts of the network, not only by defining the lattice parameters, but also by selecting an appropriate in-plane field angle. This is well known from antidot lattices, e.g., Refs.~[\onlinecite{Neusser_APL2008,Ding_JAP2011,Tacchi_2010,Gubbiotti_APL2015}], but has not been shown for artificial spin ices in general, and arrays of nanodisks in particular, until now.

\begin{figure}[t]
\centering
\includegraphics[width=0.5\textwidth]{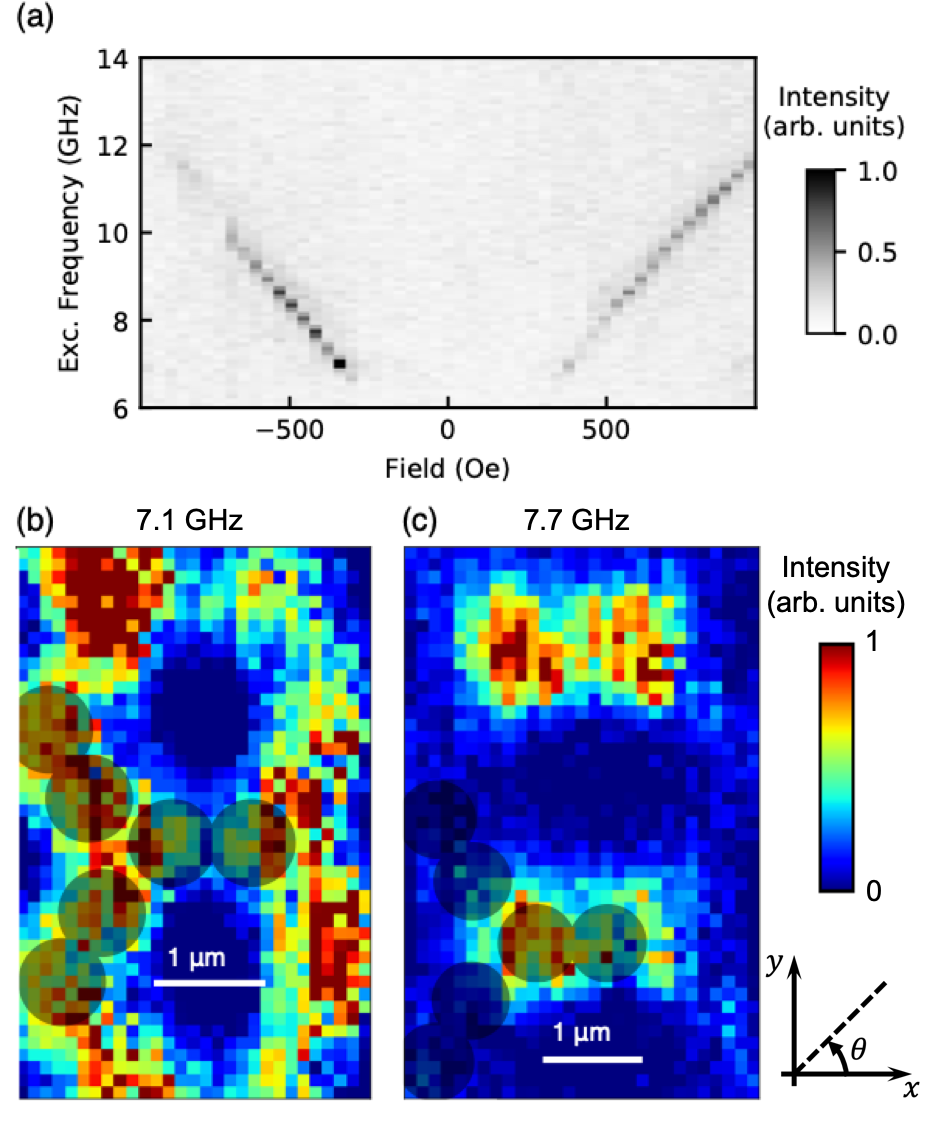}
\caption{(a) $\mu$-BLS intensity as a function of field and excitation frequency. Two-dimensional $\mu$-BLS intensity measurements at a magnetic field of $-500$~Oe and excitation frequencies of (b) 7.1~GHz and (c) 7.7~GHz. The position of one honeycomb disk vertex is superimposed in gray.}
\label{fig 7}
\end{figure}

In the following, we present experimental results obtained by $\mu$-BLS that confirm the presence of well defined \textit{channels} of spin dynamics in our structures as suggested by the simulated 2D dynamics maps. 
Our $\mu$-BLS setup allows us to apply an external magnetic field only along the CPW guide axis ($\theta=0^\circ$). A constant microwave signal of +22~dBm is applied at varying frequencies. In a first step, we record the BLS spectral response as a function of the external magnetic field and excitation frequencies for a fixed laser position on the sample, see Fig.~\ref{fig 7}(a). Although we are not able to resolve the detailed mode structure we found in FMR, overall the BLS spectrum is in agreement with the FMR results shown in Fig.~\ref{fig 2}(a). For the 2D BLS imaging, a fixed magnetic field of $-500$~Oe is chosen. At this fixed magnetic field we observe two distinct peaks in the BLS spectra for microwave excitation frequencies of 7.1~GHz and 7.7~GHz. In order to find the spatial distribution of the precessional modes, the sample was scanned at those frequencies by rastering the laser over the sample. The stage was moved in 100~nm steps, recording the BLS spectrum at each position. 
The results of the 2D scans are shown in Figs.~\ref{fig 7}(b) and (c) for the excitation frequencies of 7.1~GHz and 7.7~GHz, respectively. Note that the spectra were normalized to the elastic BLS peak to compensate for any drifts possibly occurring during the course of the measurements.

The 2D BLS images show a high-intensity signal [represented by the red color in Fig.~\ref{fig 7}] localized in different nanodisks in the lattice depending on the excitation frequency. For instance, when the excitation frequency is 7.1~GHz, the disks showing a higher intensity are the ones oriented at $\pm 120 ^\circ$ with respect to the external magnetic field, Fig.~\ref{fig 7}(b). However, when the excitation frequency is 7.7~GHz, a higher BLS intensity is observed in the disks that are oriented along the external magnetic field direction, Fig.~\ref{fig 7}(c). This is in very good agreement with our micromagnetic simulations of a single vertex at $\theta=0^\circ$; compare Fig.~\ref{fig 3}(f) with Fig.~\ref{fig 7}(b) and Fig.~\ref{fig 3}(i) with Fig.~\ref{fig 7}(c). In both the simulations and in the spatially-resolved BLS measurements, we observe that the lower-frequency precession is localized in the disks oriented at $\pm 120^\circ$, and that the higher-frequency precession is localized in the disks oriented along the magnetic field. However, our BLS measurements are unable to resolve the fine structure of the magnetization dynamics indicated by antinodes observed in the simulations at higher frequencies [Figs.~\ref{fig 3}(i) and (j)].


Previous wavevector-resolved BLS studies in ASI lattices, alongside micromagnetic simulations, gave some insights in the spatial distribution of the precession modes, in particular for a square ASI \cite{Li_JPD2016} and an anti-square ASI, consisting of an extended magnetic film with empty islands positioned in a square lattice \cite{Mamica:2018ep}. Together with other recent results \cite{bhat2019direct}, our measurements provide now direct evidence of localized magnetization dynamics in a particular ASI lattice by using spatially-resolved BLS.



\section{Conclusion}
In summary, we performed detailed experimental and theoretical characterizations of the angular-dependent spin dynamics in a new type of artificial spin-ice lattices, an array of ferromagnetic nanodisks arranged on a honeycomb lattice. Using a combination of broadband ferromagnetic resonance spectroscopy and two-dimensional dynamic micromagnetic simulations, we showed that the mode spectra at different in-plane angles are strongly affected by the microstate of the network. Different subsections of the lattice predominantly contribute to the high-frequency response of the array and the exact spatial location of the dynamics can be controlled by the excitation frequency, as well as by the in-plane field angle. Furthermore, we find indications that nucleation and annihilation of vortex-like magnetization configurations in the low-field range affect the dynamics. This is different from clusters of ferromagnetic nano-ellipses, which are typically the building blocks of artificial spin ice. Our two-dimensional micromagnetic simulations are further confirmed by optical characterizations using micro-focused Brillouin light scattering, where a good agreement is found. Our work opens up new perspectives for designing magnonic devices that combine geometric frustration in a gyrotropic vortex crystal at low frequencies with a magnonic crystal at high frequencies. While beyond the scope of the current work, this could be achieved by systematic studies of the dimensions and thickness of the disks, as well as their separation and arrangement in the network.

\section*{Acknowledgements}

Work at Delaware, including FMR measurements, micromagnetic simulations, and data analysis was supported by the U.S. Department of Energy, Office of Basic Energy Sciences, Division of Materials Sciences and Engineering under Award {DE-SC0020308}. Work at Northwestern, including experimental design, was supported under NSF Grant No. DMR 1507058. Device fabrication and thin film deposition were carried out at Argonne and supported by the U.S. Department of Energy (DOE), Office of Science, Materials Science and Engineering Division. Lithography was carried out at the Center for Nanoscale Materials, an Office of Science user facility, which is supported by DOE, Office of Science, Basic Energy Science under Contract No. DE-AC02-06CH11357.

\end{document}